\documentstyle[preprint,aps]{revtex}
\begin{document}
\draft
\begin{title}
  {\Large \bf Surface Dynamics of a Freely Standing Smectic-A Film}
\end{title}

\author{Hsuan-Yi Chen and  David Jasnow}

\address{Department of Physics and Astronomy, University of Pittsburgh, 
  Pittsburgh, PA 15260, U.S.A.}

\date{\today}
\maketitle
\begin{abstract}
  A theoretical analysis of surface fluctuations of 
  a freely standing thermotropic smectic-A liquid crystal film is provided, 
  including the effects of viscous hydrodynamics. We find two 
  surface dynamic  modes (undulation and peristaltic). 
  For long wavelengths and small frequencies in a thin
  film, the undulation mode is the dominant mode.  Permeation enters the
  theory only through the boundary conditions.  
  The resulting power spectrum is compared with existing experiments.
  It is also shown that feasible light scattering experiments on a freely 
  standing smectic-A film can reveal
  viscosity and elastic coefficients through the structure of 
  the power spectrum of both the undulation and peristaltic modes.
  \\[1ex]

\end{abstract}

\narrowtext

Recent experiments on dynamics and instabilities of
soap~\cite{soap-films} and smectic-A films~\cite{Wu} have raised
interest in the general question of dynamics of freely-standing films.
The interactions between bulk elasticity and surface tension make
freely-standing smectic-A films (FSSF) suitable systems for studying
finite-size and surface effects; hence, fluctuations in FSSF are
an important subject for both theoretical and experimental
study\cite{Holyst}.  However, during the past decade only static
properties of these systems have been considered.  To our knowledge,
the only experimental work on dynamic light scattering of
FSSF\cite{Joosten} was carried out without systematic theoretical
analysis.  Hence a theoretical investigation on the dynamics of
smectic-A liquid crystals in the presence of free surfaces can help us
to gain deeper insight to the physics of FSSF and to provide a basis
for understanding new experiments.

At free surfaces, smectic layers always orient parallel to the
smectic/air interface\cite{Fournier}.  When the system is driven out
of equilibrium, surface and bulk elasticity and hydrodynamic effects
give the film very complicated dynamical properties.  In this paper we
give the first theoretical calculation of the power spectrum of the
surface fluctuations by studying the linearized hydrodynamic equations
developed by Martin {\it et al.}  \cite{Martin}.  We show that for
thin films, the long wavelength surface fluctuations have two modes.
The undulation mode is the dominant mode at low frequency and long
wavelength.  The peristaltic mode is expected to be very small.  Only
when the thickness of the film becomes very large does the peristaltic
mode become comparable to the undulation mode. Permeation processes
are important within the boundary layer~\cite{DeGennes1,Orsay}
in order that the proper boundary conditions are satisfied. However,
our calculation shows that the power spectrum of the surface fluctuations 
are independent of the permeation constant. We also show that the
earlier experimental
study of  dynamic light scattering of FSSF\cite{Joosten} was performed in the 
limit of a very thin film such that the power spectrum has the same form as 
that of a soap film, {\it i.e.},  bulk elasticity makes no contribution 
to the power spectrum.  The scaling
relations suggested in Ref.~\cite{Joosten} for the peak position
are not valid in general, especially when the thickness of the 
film is increased,
thus allowing the bulk 
elasticity of the smectic material to play a role in the dynamics.  
For a reasonably
thick film, the power spectrum can have, in addition 
to a single undulation peak, 
some additional structure which reveals the interaction between the 
two free surfaces and the contribution from bulk properties.
We suggest that more light scattering experiments on a FSSF will be great 
help for an understanding of the interplay of the surface and bulk 
properties in this layered system.  

In the ground state, 
an ideal smectic-A phase consists in a uniform piling of planar, parallel and 
equidistant layers of molecular thickness.  We use a continuum description and
take the average layer normals 
parallel to the $z$-axis. When small fluctuations are present, 
the bulk elastic free energy of smectic-A phase is given by\cite{DeGennes1}
\begin{eqnarray}
    F=\int d^3 {\bf r} \frac{1}{2}
        \left\{B \left(\frac{\partial u}{\partial z}\right)^2 + 
        K_1 \left(\frac{\partial ^2 u}{\partial x^2} 
        + \frac{\partial ^2 u}{\partial y^2}\right)^2 \right\}
\end{eqnarray}
where $u({\bf r},t)$ is the layer displacement from the equilibrium 
position at position {\bf r} at time $t$, $B$ and $K_1$ are, respectively, the
the layer compression and undulation elastic moduli.  The characteristic length 
$\lambda \equiv \sqrt{K_1/B}$ is typically of the order of the layer spacing
($ \sim 10 ^{-7}$cm)\cite{DeGennes1}.

    The equations for viscous flow in smectic A liquid crystals 
were first written down by Martin {\it et al}\cite{Martin}.  In the 
absence of topological defects, the system satisfies the equations of 
motion in bulk \begin{eqnarray}
    \rho \frac{\partial v_i}{\partial t}=-\partial_i p
+ \partial_j  \sigma' _{ij} + h  \ \delta _{iz}
\label{eq:NS}
\end{eqnarray}
and 
\begin{eqnarray}
    \frac{\partial u}{\partial t}=v_z+\zeta_p \ h \ \ ,
\end{eqnarray}
where 
$\sigma'$ is the viscous stress tensor, $\zeta_p$ is the permeation constant
and $h$ is defined by 
\begin{eqnarray}
   h \equiv \partial_i \left( \frac{\delta F}{\delta \partial_i u}\right).
\end{eqnarray}
In (\ref{eq:NS}) we sum on repeated indices, and 
$\partial_j  = \partial / \partial x_j.$
An important length scale associated with permeation is the boundary layer
$\delta$.  Within this distance to the boundary, permeation takes place to 
satisfy proper boundary conditions of the system under consideration
\cite{DeGennes1,Orsay,note}. 
We will show that the boundary layer is associated with the
``permeation modes''.  The boundary conditions 
for free surfaces are discussed below.

    We consider a freely standing Smectic-A liquid crystal 
film with film normal
in the $z$-direction.  As noted, in equilibrium, the layers are parallel to the free 
surfaces.  When the surfaces are perturbed by an external force, 
elastic forces and dissipative effects
act to drive the system back to
the equilibrium configuration.  In this letter we study such a system by 
including the hydrodynamics in the film.  

    The geometry of the film is shown in Fig.~1; the film extends
from $z=-d$ to $z=0$ but is otherwise 
without boundaries in the $x$, $y$ directions.  
We consider surface light scattering experiments 
with momentum transfer vector {\bf q} in 
the $x$-direction; we further
assume translational invariance in the $y$-direction.  
The displacement of the upper and lower free surfaces from their 
equilibrium values are described by two functions 
$\zeta^{+}(x)$ and $\zeta^{-}(x)$.  

We look for surface wave solutions with the form
\begin{eqnarray}
    v_z=\sum _k \left\{A^{+}_k e^{S_k qz} +A^{-}_k e^{-S_k q(z-d)} \right\}
        e^{iwt+iqx}
\end{eqnarray}
where Re$(S_k) \geq 0$.  The $+$ ($-$) modes are the 
upper (lower) surface modes
which have their maximum amplitudes at $z=0$ ($z=-d$).

Two combinations of viscosities $\eta'\ \rm and\ \eta_3$~\cite{note}
and a length scale $\kappa^{-1} \equiv \sqrt{\zeta_p \eta_3} \sim 10^{-7}$cm
\cite{DeGennes1,Orsay,note}
enter the analysis. 
A dimensionless frequency is conveniently defined as
$ \Omega = - \frac{i \omega \rho}{\eta_3 q^2}$. 
The characteristic frequency $K_1 q^2 / \eta_3$ for the decay rate of the bulk 
undulation mode, reduced by $\eta_3 q^2 /\rho$, yields a dimensionless 
frequency 
  $ \mu = \frac{K_1 \rho}{\eta _3 ^2}.$
We assume incompressibility (the motion of the fluid is slow compared to the
propagation of sound), long wavelength ,{\it i.e.}, 
$(\kappa /q)^2 \gg 1, \ \  \lambda q \ll 1 $. 
We consider low frequency satisfying conditions $\frac{\Omega}{\mu}
(\lambda q)^2 \ll 1, \ \frac{\Omega^2}{\mu}(\lambda q)^2 \ll 1$.  This means
we consider a regime in which the velocity of the surface wave is slow 
compared to the typical velocity of ``second sound", $\sqrt{B/\rho}$
~\cite{DeGennes1}. 
Furthermore, since then the frequency cannot be as large 
as $K_1 \lambda ^{-2} /\eta_3$, permeation cannot have a significant 
contribution to the bulk undulation mode.~\cite{DeGennes1} 
In this frequency and wavenumber domain
we find three spatially decaying modes in the z-direction.    
One of them has long  spatial relaxation length and satisfies
\begin{eqnarray}
    S_3^2 &=&(\lambda q)^2 \ 
              \left[\frac{\left(1- \frac{\Omega}{\mu}(1- \Omega ) \right)}
                         {1+\frac{\Omega}{\mu} \ (\lambda q)^2
                              \left(\Omega - \frac{\eta '}{\eta_3} - 2 \right)}
              \right ] \\
          &\equiv& \left( \lambda q \right)^2[f(\Omega ,q)]^2. \label{f}
\end{eqnarray}                              
The other two modes ($S_1$ and $S_2$) are large compared to unity 
and satisfy
\begin{eqnarray}
    S^4 
    + \frac{\Omega}{\mu} \ (\kappa \lambda)^2 S^2  
    + \left(\frac{\kappa}{q}\right)^2
    =0.
\end{eqnarray}                     

In typical light scattering  experiments 
$q \sim 10^2 - 10^4$ cm$^{-1}$, $\omega \leq 10^8$ rad/s.
For typical materials  $\eta _3 \sim 1\ \rm p$ 
so that the long
wavelength, low frequency conditions above are satisfied.
Notice that when the denominator of the function $f(\Omega,q)$ defined above
is significantly different from unity, we cannot separate the solutions 
for $S_1, \
S_2, \ S_3$ in the way mentioned above.
The mode with relaxation $S_3$ is related to  
the dynamic generalization of the penetration length of layer 
distortion in the bulk due to surface undulation.~\cite{DeGennes1,Rapini} 
Following the terminology of Rapini,~\cite{Rapini}
from now on we refer to the $S_3$-mode as the {\em elastic mode.}
Since $S_1$, and $S_2$ depend on the permeation constant, $\zeta_p$,
they are called {\em permeation modes.}
For 
$e^{S_1q d}, \ e^{S_2qd} \gg 1$, {\it i.e.},
$d \gg (S_1 q)^{-1}, \ (S_2 q)^{-1} 
\sim \delta$,
the film thickness is large compared to the boundary layer $\delta$ 
(identified as the exponential decay length), and
the permeation modes relax in the bulk.  

   The surface displacements $\zeta ^+, \ \zeta ^-$ 
satisfy the boundary condition for the normal component of the velocity in 
linear theory
\begin{eqnarray}
   \frac{\partial \zeta ^{+(-)}}{\partial t} = v_z |_{z=0(-d)}.
\end{eqnarray}    
The other boundary conditions can be understood from the covariant elasticity
theory of smectic-A developed by Kl\'{e}man and Perodi \cite{kleman}.
For free surfaces the normal components of the stress tensor as
 well as the
normal component of the permeation force \cite{permeationforce} 
should vanish.  
The condition that $\sigma _{xz} =0$ on the free surfaces reveals 
$A_k ^{+(-)} \ll A_3 ^{+(-)}$ for $k = 1, \ 2$, 
which indicates that the permeation modes have 
negligible contributions to the dynamics of the system.  
The condition that $\sigma _{zz} =0$ yields
the following relation between the surface displacements 
and the external forces $P_{ext} ^+, \ P_{ext} ^-$ which
are assumed to act on the free 
surfaces,
\begin{eqnarray}
    P_{ext} ^{+(-)} = \left[ p-\sigma _{zz}' + B \frac{\partial u}{\partial z}
    + \alpha \frac{\partial ^2 \zeta^{+(-)} }{\partial x^2} \right] _{z=0(-d)},
 \label{stress}   
\end{eqnarray}
where $\alpha $ is the air-film surface tension.
The normal component of the permeation force, in the system under 
consideration, is 
\begin{eqnarray}
   B \frac{\partial u}{\partial z} = 0
 \label{u}  
\end{eqnarray}
at $z=0, \ -d$ for all $x$.
The linear response function $X$ connects the surface displacements with the 
external forces through
\begin{eqnarray}
    {\mbox {\boldmath$\zeta$}} (q,w)= - {\bf X}(q,w) {\bf P}_{ext}(q,w) {\it A}
\ \ ,
\label{res}   
\end{eqnarray}
where {\it A} is the surface area, {\boldmath$\zeta$}, ${\bf P}_{ext}$ 
are vectors for 
surface displacement and external force respectively, e.g.,
{\boldmath$\zeta$}$= (\zeta^+, \ \zeta^-)$;
${\bf X}$ is a $2\times 2$ matrix for the response function.

Putting Eqns.~(\ref{u}),(\ref{stress}), and  (\ref{res}) together our 
calculation leads to two 
surface dynamic modes; they are, respectively, the undulation mode 
with amplitude 
\begin{eqnarray}
\zeta ^U= \frac{1}{2} \left(\zeta ^+ + \zeta ^- \right)
\end{eqnarray}
and the peristaltic mode with amplitude 
\begin{eqnarray}
\zeta ^P= \frac{1}{2} \left(\zeta ^+ - \zeta^- \right).
\end{eqnarray}  
The response functions are in turn given by
\begin{eqnarray}
X^{U} = \frac{1}{2 \alpha q^2 {\it A}} \ \ 
        \frac{1}{1 + g \ f(\Omega, q) \ {\rm tanh}
         \left(\frac{\lambda q^2 d}{2} f(\Omega ,q) \right) }\ \ ,
\label{undu}                       
\end{eqnarray}   
and 
\begin{eqnarray}
    X^{P} =\frac{1}{2 \alpha q^2 {\it A}} \ \  
        \frac{1}{1 + g \ f(\Omega, q) 
        \ {\rm coth}\left(\frac{\lambda q^2 d}{2} f(\Omega, q) 
                        \right) 
                        } \ \ ,
\label{per}                        
\end{eqnarray}       
where $g\equiv \sqrt{BK_1}/\alpha$ measures the relative importance of bulk
elasticity and surface tension.  When the 
contribution from bulk elasticity and viscosity dominates, terms involving 
$g$ are dominant.
These equations are the central result of our
calculation.

In the limit of infinite thickness, 
$d\rightarrow \infty$, the response functions of both modes become the same, 
indicating that the interactions between the two surfaces vanishes for large 
thickness.  For 
low frequency ($\Omega \ll 100$) and infinite thickness, the result 
agrees with 
the earlier calculation
by Rapini\cite{Rapini} for an infinitely thick smectic-A material with a free
surface.     

    We now introduce a natural frequency of surface motion as 
$\omega_0 ^2=\frac{2\alpha}{\rho d}q^2$, and dissipation coefficient
$\gamma=\frac{\eta_3 q^2}{\rho}$.
In the range of the only available experiment\cite{Joosten}
(thin film, $d\sim 100$ nm, small wavenumber, 
$q\sim 10^4 \ {\rm cm}^{-1}$), $\lambda q^2d \ f(\Omega,q)/2 \ll 1$.  
The response
function is approximately
\begin{eqnarray}
X^U &\sim& \frac{1}{ 1 + \frac{1}{2}g \ [f(\Omega, q)]^2 \lambda q^2 d } 
                                                            \nonumber \\\
    &\sim& \frac{1}
           {(\omega_0 ^2 - \omega^2) +i\gamma \omega} \ ,
\end{eqnarray}
which is the same as the response function of a damped driven simple harmonic
oscillator.  The peak position in the range of weak damping 
($\frac{\eta_3 q^2}{\rho}\omega \ll \sqrt{\frac{2\alpha}{\rho d}} q$ ) is 
\begin{eqnarray}
    \omega= q\sqrt{\frac{2\alpha}{\rho d}}
\end{eqnarray}
This is the same as for  a soap film and gives the scaling relation which the 
experimental data satisfy.  In general we may not have a system with 
weak damping, and the peak position should be slightly modified.  
On the other hand,
the width of the undulation mode peak ($\eta_3 q^2 /\rho$) does provide  
 information about the viscosity $\eta_3$.
 
The peristaltic mode changes the layer spacing much more significantly than the
undulation mode; hence we expect that the peak in the peristaltic
power spectrum occurs when
the term proportional to $g$ in Eq. (\ref{per}) dominates. Also in the range that
we are interested in, the relation $\Omega / \mu \gg 1$ is valid.  A 
straightforward calculation leads to an estimate of the peak position of 
the peristaltic mode :
\begin{eqnarray}
        \lambda q^2 d \ f_I(\Omega,q) \approx \pi
\end{eqnarray}
where $f_I(\Omega,q)$ is the imaginary part of the function $f$ we have defined
via Eq. (\ref{f}).
In this approximation the surface tension plays no role in 
peristaltic mode, the peak position is essentially independent of $g$ and has
the form
\begin{eqnarray}
    \Omega \approx -i \ \frac{2\mu \pi^2}
                                 {\lambda q^2 d 
                                  \sqrt{\lambda^2 q^4 d^2 + 4 \mu \pi^2}  }.
\end{eqnarray}
For typical material parameters the above approximation yields the peristaltic
peak within 5\%.                
From this relation we can estimate the bulk elastic coefficients $K_1$ and
$B$ by fitting the value $\mu$ and $\lambda$ from the peristaltic peak.

One may ask whether it is possible to observe the peristaltic mode and
the special features of a smectic-A liquid crystal in a free-standing
film experiment. Figure 2 shows the power spectrum of one free surface
,{\it i.e}, $\zeta^U + \zeta^P$,
for a typical choice of parameters with increasing film thickness.  As
the film gets thicker we observe that a peak develops in the higher
frequency range.  This peak is due to the peristaltic mode.  
When $L=q d \ge 12, \ i \ \Omega \ge 3$ there is some extra structure arising
from sources other 
than the peristaltic mode.  Figure 3 shows the power spectrum for 
the same choice of
parameters with $q d=15$.  The contribution from both undulation
and peristaltic modes are plotted in long and short dashed lines,
respectively.  We find that the power spectrum of the peristaltic mode
is about one order of magnitude smaller than the contribution
of the undulation mode.
Notice that there is a second peak of the undulation mode in the high
frequency region.  It comes from the oscillating behavior
of the hyperbolic tangent function when the argument ( $\lambda q^2 d
\ f(\Omega,q)/2$ ) is complex. 
This extra structure in the power spectrum at high $\Omega$ is a
result of the interplay of bulk elasticity and the existence
of the two free surfaces; it is a special feature of a structured
fluid. 
Hence we conclude that for reasonable choices of 
material parameters and for an experiment with 
typical dynamic range, it should be possible to observe the
peristaltic mode.  
Combinations of material parameters can be extrapolated from the 
shape of the undulation peak and the peristaltic peak.
However, the detailed shape of the power spectrum is sensitive to the specific
material parameters used in a laboratory experiment.

In conclusion, we have derived the power spectrum of a
freely standing smectic-A film within linear hydrodynamics and
assuming the absence of topological defects. 
The dynamics of this
system are dominated by the elastic mode and the permeation constant does not 
show up in the power spectrum of the surfaces.  
The permeation process is important
near the surfaces and helps the system to satisfy 
proper boundary conditions.   
When the thickness of the film is small enough, bulk elasticity does not
contribute to the undulation mode, and the peristaltic mode is not
observable.  However, for a reasonably thick film the power spectrum does show
the interplay of surface tension and bulk elasticity.  The extra structure 
in the power spectrum is a special feature due to the existence of  two
free surfaces and the contribution from the bulk elasticity.  We suggest that 
further experimental work on FSSF over a wider range of film thickness and
wavenumber can observe these interesting features.

We thank Prof.~X-l. Wu and D. Dash for very helpful discussions, and
Prof. Takao Ohta for his interest and assistance.
H.Y.C. acknowledges 
financial support from University of Pittsburgh as an 
Andrew Mellon Predoctoral
Fellow.  D.J. is grateful for the support of the NSF under 
DMR9297135.

\newpage
{\large \bf Figure Caption}

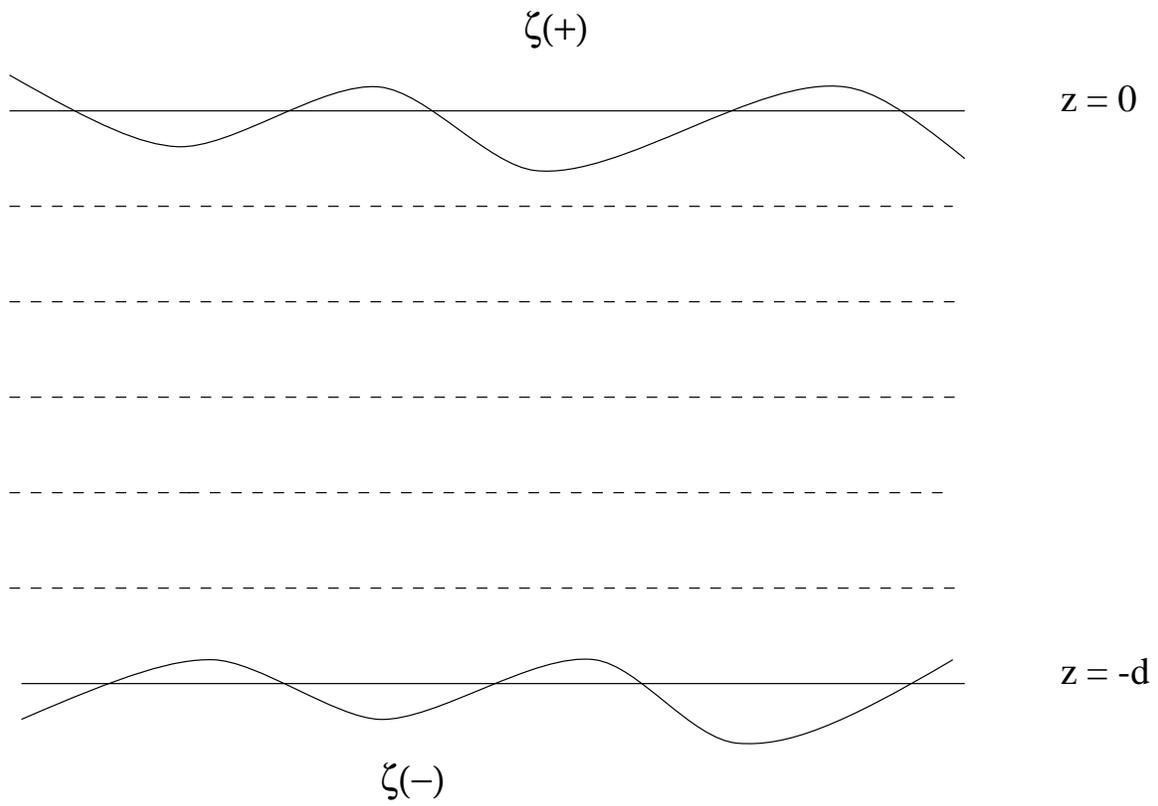
Figure 1. Schematic of a freely standing smectic-A film of thickness d.  
          
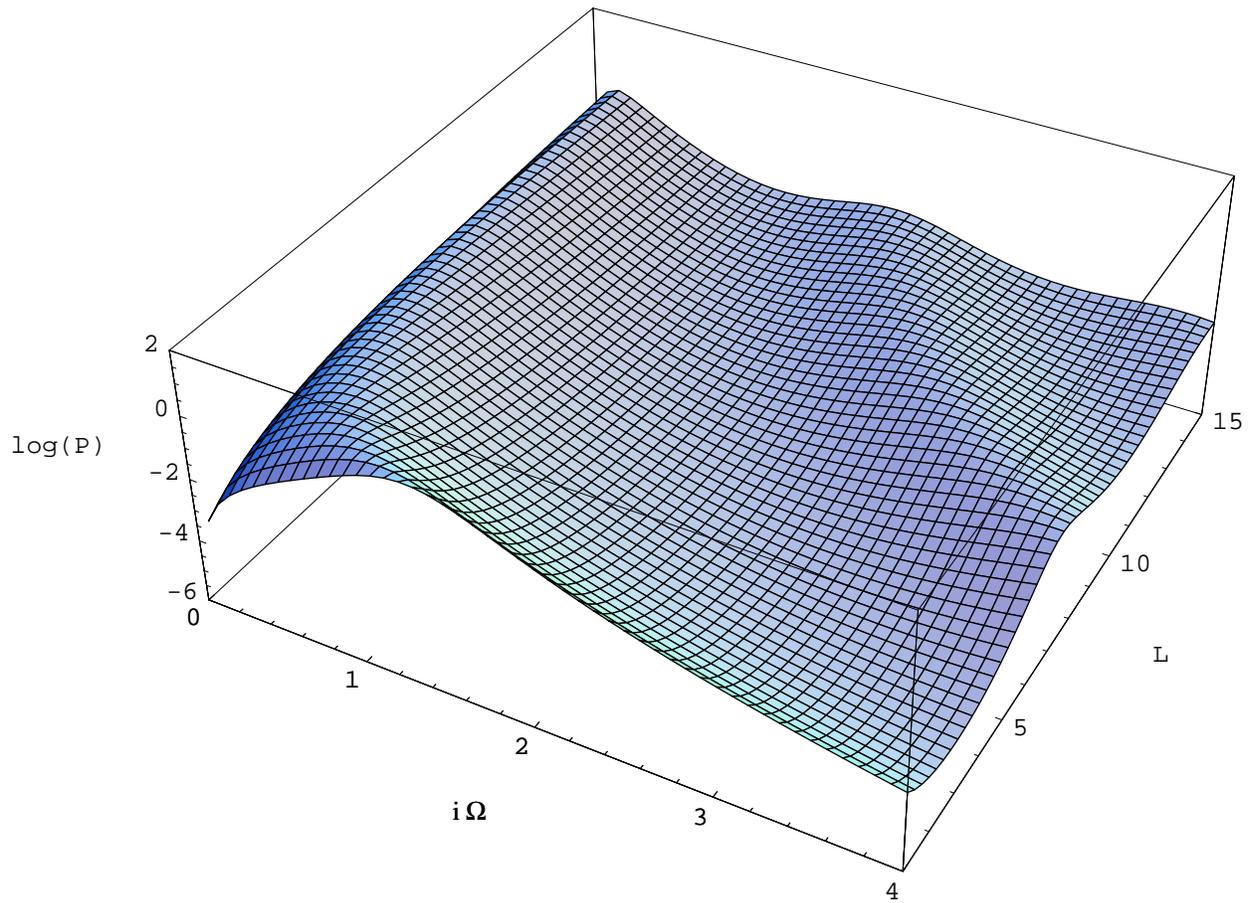
Figure 2. The power spectrum (natural logarithm)
          for a typical choice of parameters with increasing
          thickness of the film:
          $\lambda q=10^{-3}$, $\mu=10^{-4}$, $g=0.1$, 
          $\eta' = \eta_3$.  
          $L=q d$ is dimensionless.
          As the thickness increases, we can easily see the peristaltic mode
          and extra structure of the undulation mode provide additional
           contribution to the power spectrum.           
                         
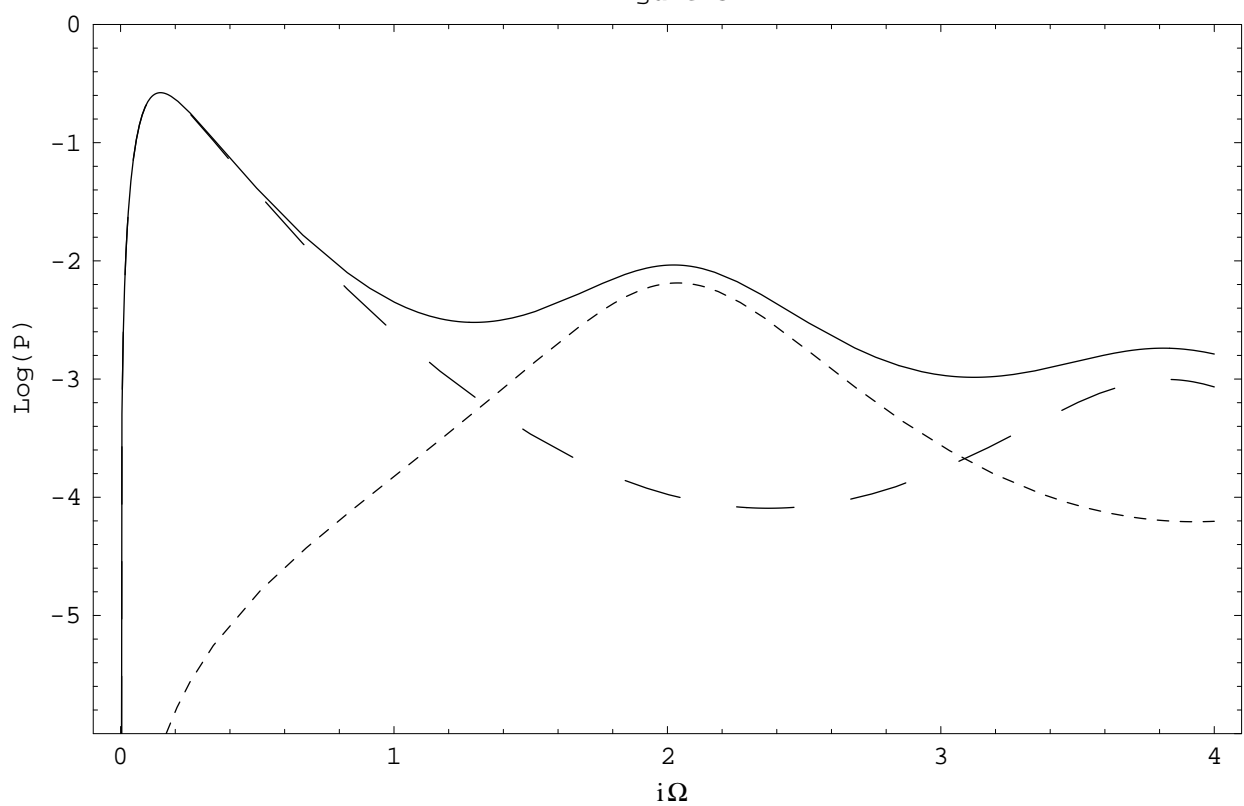
Figure 3. The power spectrum (natural logarithm) 
          for a typical choice of parameters:
          $\lambda q=10^{-3}$, $q d=15.0$, $\mu=10^{-4}$, $g=0.1$, 
          $\eta' = \eta_3$. 
          The long dashed line is the undulation mode, short dashed line
          is the peristaltic mode, solid line is the total power spectrum.
          Notice that there is a small peak in the undulation mode at 
          $ i \  \Omega \sim 4$.


\begin{references}
\bibitem{note} The viscosity coefficient $\eta_3$ is defined in 
Ref.~\cite{Martin}. In their notation,
$\eta '=\eta_1 + \eta_2 - 4 \eta_3 - 2 \eta_5 + \eta_4$.
\bibitem{permeationforce} In Kl\'{e}man's notation, the permeation 
force on the surface is the vector {\bf S} (defined in 
Ref.~\cite{kleman}), which happens to be 
$B \frac{\partial u}{\partial z}$  when the normal of
the surface is in the $z-$direction.  The torque acting
on the surface is always zero in the system under
consideration.
\end{references}
\end{document}